\documentclass[fleqn,twoside]{article}
\usepackage{espcrc2}

\usepackage{graphicx}
\usepackage[figuresright]{rotating}

\newcommand{\AmS}{{\protect\the\textfont2
  A\kern-.1667em\lower.5ex\hbox{M}\kern-.125emS}}

\newcommand{\FLpcac}{ F_L^{\rm PCAC} }
\newcommand{\FLac}{ F_L^{\rm A} }

\newcommand{\FLvc}{ F_L^{\rm V} }
\newcommand{\FTvc}{ F_T^{\rm V} }

\title{Cross-section measurements in the NOMAD experiment}  

\author{R. Petti\address[MCSD]{
        CERN, CH-1211 Gen\'eve 23, Switzerland}\address[MCSD]{  
        University of South Carolina, Columbia SC 29208, USA}  
        \thanks{For the NOMAD collaboration}\thanks{E-mail address: Roberto.Petti@cern.ch}}
       
\begin{document}

\begin{abstract}
The NOMAD experiment collected valuable neutrino data samples,
matching both the large statistics of massive calorimeters and
the reconstruction quality of bubble chambers. This paper describes 
the recent measurements of neutrino cross-sections on carbon target. 
The approach followed for cross-section modeling is also explained.  
\vspace{1pc}
\end{abstract}

\maketitle

\section{Introduction}
\label{sec:intro}

The NOMAD experiment was designed to search for $\nu_{\tau}$
appearance from neutrino oscil\-la\-tions in the CERN wide-band
neutrino beam produced by the 450 GeV proton synchrotron.
The single-particle reconstruction and lepton identification capability
of the NOMAD detector allowed the search for $\nu_{\tau}$ appearance in most
of the leptonic and hadronic $\tau$ decay channels~\cite{nmnt} and
also to look for $\nu_{\mu} \rightarrow \nu_{e}$ oscillations~\cite{nmne}.
No evidence for oscillations was found.

The high resolution and reconstruction quality makes the NOMAD 
data samples a valuable resource for the neutrino physics community.  
The recent measurements of cross-sections and particle production 
would help to clarify our understanding of neutrino interactions  
with nuclei at intermediate energies. In addition, it must be noted 
the precise tracking in a light target provides a powerful tool to 
tune and validate Monte Carlo simulation programs for future experiments. 
This analysis activity can benefit from the beam and detector
studies performed for the oscillation searches.

\section{Detector and data samples}
\label{sec:detdata}

The NOMAD detector is described
in detail in Ref.~\cite{NOMADNIM}. Inside
a 0.4 T magnetic field there is an active target consisting of
drift chambers (DC)~\cite{DC} with a fiducial mass of about 2.7 tons
and a low average density (0.1 g/cm$^3$). The main target, 405 cm
long and corresponding to about one radiation length,
is followed by a transition radiation detector (TRD)~\cite{TRD}
for electron identification, a preshower detector (PRS), and
a high resolution lead-glass electromagnetic
calorimeter (ECAL)~\cite{ECAL}. A hadron calorimeter (HCAL) and
two stations of drift chambers for muon detection are located
just after the downstream part of the magnet coil. An iron-scintillator
sampling calorimeter with a fiducial mass of about 17$t$ (FCAL) is
located upstream of the central part of the NOMAD target.
The detector is designed to identify leptons and to measure
muons, pions, electrons and photons with comparable resolutions.
Momenta are measured in the DC with a resolution:
$$
\frac{\sigma_p}{p}\simeq \frac{0.05}{\sqrt{L[m]}}\oplus
\frac{0.008 \times p[GeV/c]}{\sqrt{L[m]^5}}
$$
where L is the track length and $p$ is the momentum.
The energy of electromagnetic showers, $E$, is measured
in the ECAL with a resolution:
$$
\frac{\sigma_E}{E}=0.01 \oplus \frac{0.032}{\sqrt{E[GeV]}}.
$$

The relative composition
of Charged Current (CC) events in NOMAD is estimated~\cite{beam} to be
$\nu_{\mu}$,:\,$\bar{\nu}_{\mu}$,:\,$\nu_e$,:\,$\bar{\nu}_e$ =
1.00\,: \,0.0227\,: \,0.0154: \,0.0016, with average
neutrino energies of 45.4, 40.8, 57.5, and 51.5 GeV, respectively.
Neutrinos are produced at an average distance of 625 m from the detector.

The NOMAD experiment collected data from 1995 to 1998. Most of the running,
for a total exposure of $5.1\times 10^{19}$ protons on target (pot), was
in neutrino mode. This resulted in three distinct data samples, according
to the different targets: $1.3\times 10^{6} \nu_{\mu}$ CC interactions
from the drift chambers (mainly carbon), $1.5\times 10^{6} \nu_{\mu}$ CC interactions
from the region of the magnet coil (mainly aluminium) located in front of the DC
and $1.2\times 10^{7} \nu_{\mu}$ CC interactions from FCAL (iron).

\section{Modeling of inelastic cross-sections}
\label{sec:model}

\subsection{Structure functions at high $Q^2$}

At large momentum transfer structure functions are described as series 
in $Q^{-2}$ on the basis of the operator product expansion of
the correlator of the weak current (twist expansion):  
\begin{eqnarray}
\label{eq:SFs}  
F_i(x,Q^2) & = & F_i^{\rm TMC}(x,Q^2)  \nonumber \\  
        & + & \frac{H_i^{(4)}(x)}{Q^2}
        + \frac{H_i^{(6)}(x)}{Q^4}
        + \cdots,
\end{eqnarray}
where $i=T,2,3$ refers to the type of the structure function,
$F_i^{\rm TMC}$ are the leading twist (LT) terms corrected for the target mass
effects and $H_i^{(t)}$ are the higher twist (HT) terms of twist $t$. The target mass 
corrections are computed using the approach of Ref.\cite{GeoPol76} (see also \cite{KP04} 
for the treatment of the thereshold problem at $x \to 1$). The calculation of the 
leading twist is performed in the NNLO approximation and we include additional 
phenomenological terms up to twist-6. We use PDFs and HT based on Ref.\cite{A02},
obtained from dedicated fits optimized at low $Q^2$~\cite{AKP06} and including additional
data from (anti)neutrinos CC (NOMAD, CHORUS~\cite{chorus-xsec} and NuTeV~\cite{nutev-xsec}) 
and charged lepton (JLab) Deep Inelastic Scattering (DIS), as well as from 
Drell-Yan production (E605 and E886).

\subsection{Low $Q^2$ structure functions}

In the low-$Q$ region (anti)neutrino cross sections are
dominated by the longitudinal structure function $F_L$ and the latter is
driven by the axial-current interactions. The structure function $F_T$ vanishes 
as $Q^2$ at low $Q^2$. This behaviour is similar to the charged lepton case and holds 
for both the vector and the axial-vector contributions. However, 
in the longitudinal channel the low-$Q$ behavior of the vector and
axial-vector parts are different.
 
The conservation of the vector current (CVC) suggests $q_\mu W_{\mu\nu}=0$ for the 
vector current part of hadronic tensor. From this condition we conclude $\FLvc$
vanishes faster than $\FTvc$ at low $Q^2$ and $\FLvc/\FTvc\sim Q^2$.
This behavior is similar to the charged-lepton case.

In contrast to the vector current, the axial-vector current is not
conserved. For low momentum transfer the divergence of the
axial-vector current is proportional to the pion field
(Partially Conserved Axial Current or PCAC)
\begin{equation}
\label{pcac}
\partial A^\pm = f_\pi m_\pi^2 \varphi^\pm.
\end{equation}
where $m_\pi$ is the pion mass and $f_\pi=0.93 m_\pi$ is the pion decay
constant and $\varphi^\pm$ is the pion field in the corresponding charge
state. We introduce explicitely a PCAC contribution to $\FLac$: 
\begin{equation}
\label{FL:ac}
\FLac = \gamma^{3} \FLpcac f_{\rm PCAC}(Q^2)    
+ \widetilde{F}_L^{\rm A}
\end{equation}
where $\gamma=(1+4x^2M^2/Q^2)^{1/2}$, $\FLpcac=f_\pi^2\sigma_\pi/\pi$ and 
$\sigma_\pi=\sigma_\pi(s,Q^2)$
is the total cross section for the scattering of a virtual pion with
momentum $q$ and the center-of-mass energy squared $s=(p+q)^2$. 
The last term $\widetilde{F}_L^{\rm A}$ is similar to $\FLvc$ and vanishes as $Q^4$.
Since the PCAC contribution is expected to vanish at high $Q^2$ we introduce a
form factor $f_{\rm PCAC}(Q^2)=(1+Q^2/M_{\rm PCAC}^2)^{-2}$, where the 
dipole form is motivated by meson dominance arguments. It is important to note 
the pion pole does not directly contribute to structure functions and hence 
the mass scale controlling the PCAC mechanism, $M_{\rm PCAC}$, cannot be 
the pion mass itself, but is rather related to higher mass states like 
$a_1$, $\rho \pi$ etc. We use the simple assumption $M_{\rm PCAC} = 
m_{a_1}$~\cite{KP06}.       

\begin{figure}[htb]
\includegraphics[width=1.00\linewidth]{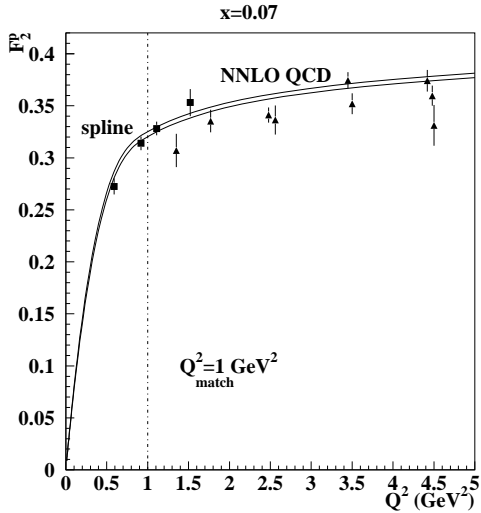}
\caption{Interpolation of structure functions in the region $0<Q^2<1~GeV^2$. 
The example given in the plot refers to $F_2$ for charged lepton scattering on 
protons at x=0.07 (see text for details).}    
\label{fig:lowQ}
\end{figure}

The structure functions $F_T$ and $\widetilde{F}_L=\FLvc + \widetilde{F}_L^{\rm A}$, 
which are vanishing for $Q^2 \to 0$ like in the charged lepton case, are parameterized 
as smooth interpolations between the high $Q^2$ regime calculated from Equation~\ref{eq:SFs} 
and the $Q^2 \to 0$ predictions derived from current conservation arguments~\cite{AKP06}\cite{M05}. 
We choose the value $Q^2_0 = 1~GeV^2$ as matching point for the twist expansion. In the 
region $0<Q^2<1~GeV^2$ we use cubic splines calculated for fixed $x$ values. 
The coefficients of such functions are fully determined by the condition   
both functions and derivatives should match with the twist expansion 
at $Q^2_0$. Figure~\ref{fig:lowQ} illustrates the interpolation procedure for 
$F_2$ on protons in charged lepton scattering.

\begin{figure}[htb]
\includegraphics[width=1.00\linewidth]{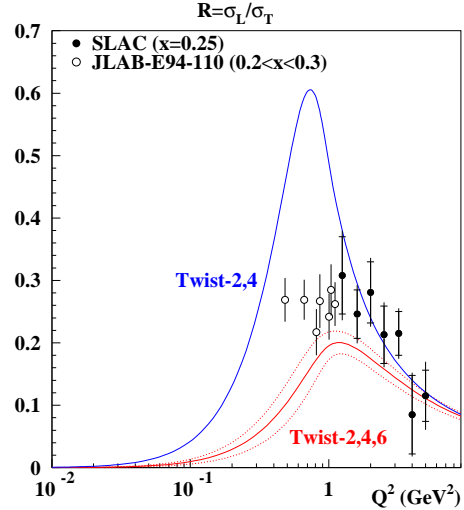}
\caption{Impact of higher twist terms on $R=\sigma_L/\sigma_T$ for charged lepton 
scattering. Data from SLAC and JLab are shown for comparison.}
\label{fig:Rcl}
\end{figure}

The different $Q^2$ dependence of various terms in Equation~\ref{eq:SFs} 
allows to disentangle higher twist contributions, which are parameterized 
as smooth functions (splines) of $x$. Our results indicate the twist-6  
term $\propto 1/Q^4$ is important in order to describe the ratio $R$ of 
longitudinal to transverse cross-sections at low $Q^2$. This is shown 
in Figure~\ref{fig:Rcl} together with SLAC and JLab data. We derive the 
ratio $R$ directly from the parameterizations of $F_L$ and $F_T$.

\begin{figure}[htb]
\includegraphics[width=1.00\linewidth]{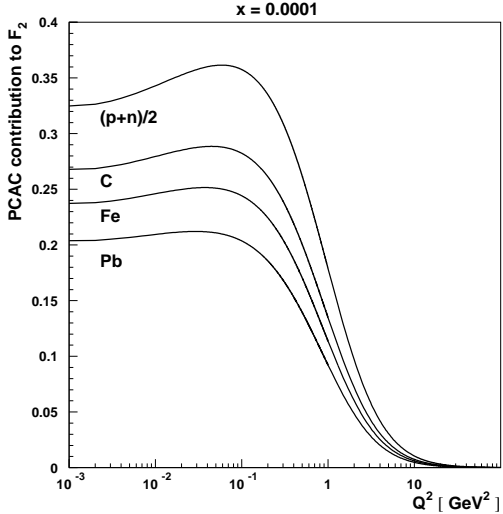}
\caption{The PCAC contribution to the neutrino structure function $F_2$ at 
$x=10^{-4}$ for different targets.}
\label{fig:F2pcac}
\end{figure}

From the relation $F_2 = (F_L+F_T)/\gamma^2$ and (\ref{FL:ac}) it
follows that the structure function $F_2$ at low $Q^2$ is dominated by the
nonvanishing $\FLpcac$ term (Figure~\ref{fig:F2pcac}). It is important to note that since $F_T\to0$ and 
$F_L\to \FLpcac$ in the limit of vanishing $Q^2$ the ratio $R=F_L/F_T$ is divergent 
for neutrino interactions. This is substantially different from the scattering of
charged leptons for which $R$ is vanishing as $Q^2$.

\begin{figure}[htb]
\includegraphics[width=1.00\linewidth]{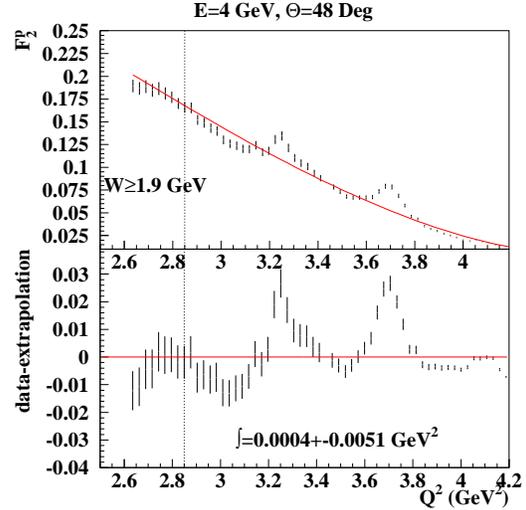}
\caption{Comparison between the extrapolation of DIS $F_2$ structure function on 
proton and the recent JLab data in the resonance region.}
\label{fig:dual}
\end{figure}

The determination of LT and HT terms is performed from all available 
data with $Q^2>0.5~GeV^2$ and $W>1.9~GeV$. It is interesting to check the 
extrapolation of DIS structure functions into the resonance region. 
The results are consistent with the duality principle, as can be seen 
from Figure~\ref{fig:dual} where the integral of the difference between 
the recent JLab resonance data and the average DIS predictions is 
consistent with zero.

\subsection{Nuclear and electroweak corrections}

A detailed calculation of nuclear corrections to structure functions is 
performed~\cite{KP04}\cite{K05}. The model takes into account a number of different effects 
including nuclear shadowing, Fermi motion and binding, nuclear pion excess and off-shell
correction to bound nucleon structure functions. The off-shell effect and the effective 
scattering amplitude describing nuclear shadowing are expressed in terms of few parameters 
which are common to all nuclei and have a clear physical interpretation. The parameters are
then extracted from statistical analysis of data from charged lepton scattering on a 
wide range of nuclear targets. 

The treatment of Fermi motion, binding, off-shell effect and nuclear pion correction 
in (anti)neutrino interactions is similar to the one in charged lepton scattering. 
The main differences are related to the impact of the axial-vector current on 
coherent nuclear effects and are more evident at low $Q^2$~\cite{K05}. 
 
One-loop electroweak effects are taken into account~\cite{ewcorr} as corrections  
to the parton distributions used in the structure function calculation.   
The initial quark mass singularities from QED corrections are subtracted 
within the $\bar{\rm MS}$ scheme since they are effectively incorporated 
in the quark density functions. Results are cross-checked with a second 
independent calculation~\cite{ewcorr2}.

\subsection{Neutrino fluxes}

Two independent approaches are used to calculate neutrino fluxes.
The first method is based on a simulation of the West Area 
Neutrino Facility~\cite{beam}. The calculation of particle production rates from the interaction 
of primary protons on $Be$ target is performed with a recent version of FLUKA~\cite{FLUKA},     
further modified to take into account the cross-sections measured by the SPY and NA20 
experiments. These particles are then propagated through the beam line taking into 
account the material and magnetic fields they traverse. Predictions are then validated 
by comparisons with NOMAD data.  

A second method relies directly upon the $\nu_{\mu}$ CC events reconstructed in the NOMAD 
detector. 
In particular, events with low energy associated to the hadronic system ($\nu< 3$ GeV) 
are analyzed to this purpose since the corresponding differential distribution $dN/d\nu$ 
is proportional to the flux up to a small correction factor ($\sim 10\%$).

\subsection{Monte Carlo simulations}

The Monte Carlo (MC) simulation of neutrino DIS interactions is based on
LEPTO 6.1~\cite{LEPTO} and JETSET 7.4~\cite{JETSET} packages, followed
by a full GEANT3~\cite{GEANT} propagation to model the detector response.
The neutrino cross-sections are parameterized according to the model 
described in the previous Sections. The simulation of resonance 
production is performed according to Ref.~\cite{RSmodel}.   

We do not include the parton shower treatment from JETSET. The reinteractions of
hadrons with surrounding nucleons in target nuclei are described
within the DPMJET~\cite{DPMJET} package. A detailed tuning of fragmentation 
parameters has been performed with hadrons reconstructed in $\nu_{\mu}$ 
CC interactions~\cite{CP06}.

\section{Measurement of inelastic cross-section}
\label{sec:dsigma2}

The present knowledge of neutrino cross-sections is rather
nonuniform. In the region $E_{\nu}>30$ GeV, where data from the large
massive calorimeters (CCFR, NuTeV) are available, the uncertainty is
about 2\%. This increases to about 20\% at lower energies, due
to the limited statistics of bubble chamber experiments.
Measurements of both the total $\sigma_{CC}^{\rm tot}$
and the differential $d\sigma^{2}_{CC}/dx dy$ cross-sections for $\nu_{\mu}$
on carbon are performed in NOMAD. One of the primary goals is to 
constrain the (anti)neutrino cross-section model and to reduce the 
corresponding systematic uncertainties. This also provides the first 
measurement of inclusive cross-section on carbon target, with  
$\langle Q^2 \rangle \sim 13$ GeV$^2$.   

The absolute normalization is obtained from the world average
cross-section on isoscalar target in the energy range $40<E_{\nu}<300$ GeV.
The measurement is performed in the kinematic region $\nu > 3$ GeV 
and $Q^2>1$ GeV$^2$. The analysis of the lower $Q^2$ region down to 
about 0.3 GeV$^2$ is currently being finalized.    

\begin{figure}[htb]
\hspace*{-0.20cm}\includegraphics[height=11.0cm,width=1.10\linewidth]{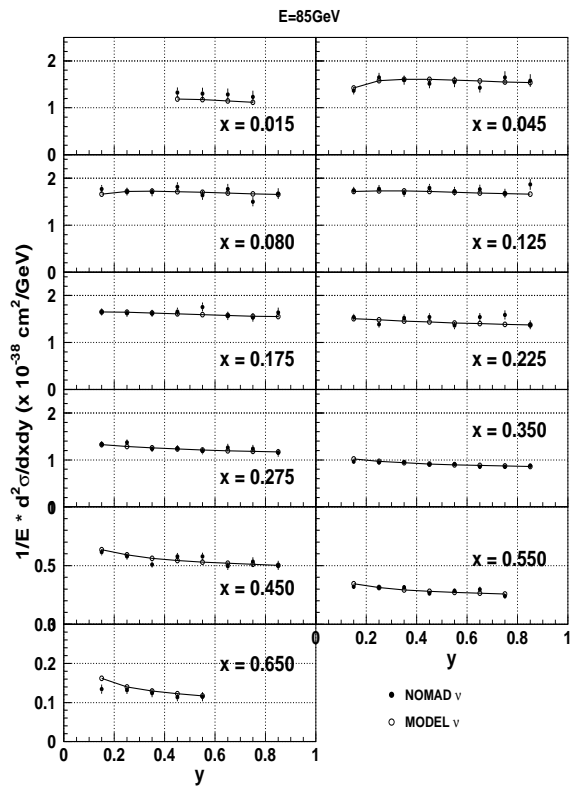}
\caption{Differential neutrino CC cross-section on carbon at $E=85$ GeV. 
The full circles show the NOMAD measurement while the curve with open circles  
is the prediction from our cross-section model.}   
\label{fig:nomad_dxy}
\end{figure}

A comparison of the measured differential cross-section on carbon
with the model predictions (Sec.~\ref{sec:model}) is illustrated in 
Figure~\ref{fig:nomad_dxy} for $E=85$ GeV, indicating a good 
agreement in the entire kinematic region. 

The NOMAD results are also consistent with the measurements performed 
by CHORUS and NuTeV on different targets~\cite{K05}. Only in the recent NuTeV data a 
slight excess is observed for $x>0.5$~\cite{K05}. The low $Q^2$ data from CHORUS 
support our treatment of the PCAC contribution to neutrino structure functions.

\section{Measurement of quasi-elastic cross-section}
\label{sec:QE}

The reconstruction and identification of the recoiling proton
track allows a measurement of the quasi-elastic cross-section
$\nu_{\mu} n \rightarrow \mu^{-} p$ on carbon in NOMAD.
A kinematic analysis based upon a three-dimensional
likelihood function is used to reject backgrounds from
DIS and resonance production. Overall, about 8000 two track events 
are selected for the cross-section determination in the 
energy range $3<E_{\nu}<100$ GeV. A complementary sample 
of about 16000 single track events is used as a cross-check of  
the reconstruction efficiency. 

In order to reduce systematic uncertainties, NOMAD is
measuring the ratio of quasi-elastic cross-section
with respect to DIS ($W^{2}>4$ GeV$^2$) processes. 
The absolute normalization is provided also in this case by
the world average cross-section on isoscalar target.  

The simulation of the quasi-elastic events is based 
upon the formalism by Llewellyn-Smith~\cite{LLS} and the     
axial form factor is parameterized in the conventional 
dipole form. Nuclear effects and the Pauli suppression 
factor are implemented following a simple Fermi gas model. 
Final state interactions are described by the DPMJET package~\cite{DPMJET}. 
The main systematic uncertainties are related to the 
understanding of nuclear corrections and are currently being  
finalized.

\end{document}